\documentclass{wiley-article}

\usepackage{siunitx}

\papertype{Original Article}
\paperfield{Journal Section}
\usepackage{xurl}
 \usepackage{setspace, amsmath, amssymb, url, lscape, subfig, algorithmic, multirow, pslatex, listings, verbatim, alltt, amsfonts, wrapfig, boxedminipage, color, cite, bookmark}
\usepackage{caption} 
\usepackage{booktabs}
\usepackage{algorithmic}
\usepackage{algorithm}
\usepackage{balance}
\newcommand{\eg}{{\it e.g., }}

\newcommand{\ie}{{\it i.e., }}
\newcommand{\name}{SMSE}
\newcommand{\Fname}{Serverless Multimedia Streaming Engine }

\newcommand{\ssp}{stream service provider}

\newcommand{\comments}[1]{}
\newcommand{\revised}[1]{{#1}}

\DeclareMathDelimiter{(}{\mathopen} {operators}{"28}{largesymbols}{"00}
\DeclareMathDelimiter{)}{\mathclose}{operators}{"29}{largesymbols}{"01}

\title{SMSE: A Serverless Platform for Multimedia Cloud Systems}

\begin{document}
\author[1\authfn{1}]{Chavit Denninnart}
\author[2\authfn{2}]{Mohsen Amini Salehi}


\affil[1]{High Performance Cloud Computing (HPCC) Laboratory, School of Computing and Informatics, University of Louisiana at Lafayette, Louisiana, 70503, USA}

\affil[2]{High Performance Cloud Computing (HPCC) Laboratory, Computer Science and Engineering Department, University of North Texas, USA}

\corraddress{Mohsen Amini Salehi, Computer Science and Engineering Department, University of North Texas, 76207, USA}
\corremail{mohsen.aminisalehi@unt.edu}


\fundinginfo{National Science Foundation (NSF) Award Numbers: 2047144 and 1940619}

\runningauthor{Denninnart and Amini Salehi}

\begin{frontmatter}
\maketitle


\begin{abstract}
Along with the rise of domain-specific computing (ASICs hardware) and domain-specific programming languages, we envision that the next step is the emergence of domain-specific cloud platforms. 
Considering multimedia streaming as one of the most trendy applications in the IT industry, the goal of this study is to develop \name, the first domain-specific serverless platform for multimedia streaming. \name~democratizes multimedia service development via enabling content providers (or even end-users) to rapidly develop their desired functionalities on their multimedia contents. Upon developing \name, the next goal of this study is to deal with its efficiency challenges and develop a function container provisioning method that can efficiently utilize cloud resources and improve the users' QoS. In particular, we develop a dynamic method that provisions durable or ephemeral containers depending on the spatiotemporal and data-dependency characteristics of the functions. Evaluating the prototype implementation of \name~under real-world settings demonstrates reducing both the containerization overhead, and the makespan time of serving multimedia functions (by up to 30\%) in compare to the function provision methods being used in the general-purpose serverless cloud systems. 

\end{abstract}


\end{frontmatter}

\section{Introduction}
\label{sec:intro}



%
The sunset of Moore's law has shifted the computing landscape from producing different forms of CPU to manufacturing systems with domain-specific co-processors, a.k.a. Application-Specific Integrated Circuits (ASICs). Google TPUs \cite{abadi2016tensorflow}, AWS Inferentia \cite{awsinferentia}, Data Processing Units (DPUs) \cite{wu2014q100}, Intel quick sync video~\cite{intelquicksync}, Xilinx FPGA video encoder \cite{twitchfpga}, along with various models of Nvidia GPU \cite{cuda} are just a few popular mentions in this drastic shift. It is anticipated that the datacenters with such ASICs machines will be the cornerstone of next generation cloud computing systems. Moreover, domain-specific programming languages and tools are emerging for popular applications (\eg machine learning, cryptography, multimedia processing, and even fluid dynamics \cite{friebel_fpga4hpc21}) to both ease the developers' job and shorten the software production cycles. 

On another front, serverless Computing and Function-as-a-Service (FaaS) are gaining popularity as a low-overhead and highly scalable cloud computing paradigm that abstracts the developers from the resource allocation and management details. Nonetheless, it is proven that a generic serverless platform cannot optimally fit all the use cases~\cite{eismann2020serverless,eismann2020review,lloyd2018serverless}. For instance, a serverless platform that is designed for cost-efficiency using ephemeral containerized functions does not fit applications with low-latency constraints \cite{shahrad2020serverless}. Moreover, applying the serverless principles on the applications that do not lend themselves to this paradigm can potentially cause a performance degradation. For instance, a data-intensive application that frequently interacts with the storage system does not match with the stateless nature of serverless functions \cite{sreekanti2020cloudburst}, and a single invocation of such function (a.k.a. \emph{task}) can lead to a considerable initialization delay. 

Within these trends and with the prevalence of serverless computing, \emph{we envision that domain-specific serverless cloud platforms are emerging for popular applications as the second generation of cloud computing systems} that can benefit both the \emph{users} and the \emph{providers}. While first generation of cloud systems aimed at mitigating the burden for system administrators, the second generation targets at alleviating the burden for programmers and solution architects via offering high-level abstractions, APIs, and services within a certain application domain \cite{patterson21}. In the one hand, users of a domain-specific serverless platform are offered specialized services with builtin QoS-awareness and application-specific programming abstractions that together can unlock new use cases. 
On the other hand, providers of a domain-specific cloud can improve their resource utilization and employ granular reusing techniques at the function level, via aggregating similar functions calls across different applications \cite{samadi2014paraprox}. With this vision in mind, \emph{we observe that there is an infrastructural gap for a platform that can fulfill the goals of a domain-specific cloud system}. It is this infrastructural gap that this paper aims to fill.

\begin{figure}[ht]
 \begin{center}
    \includegraphics[width=0.7\textwidth]{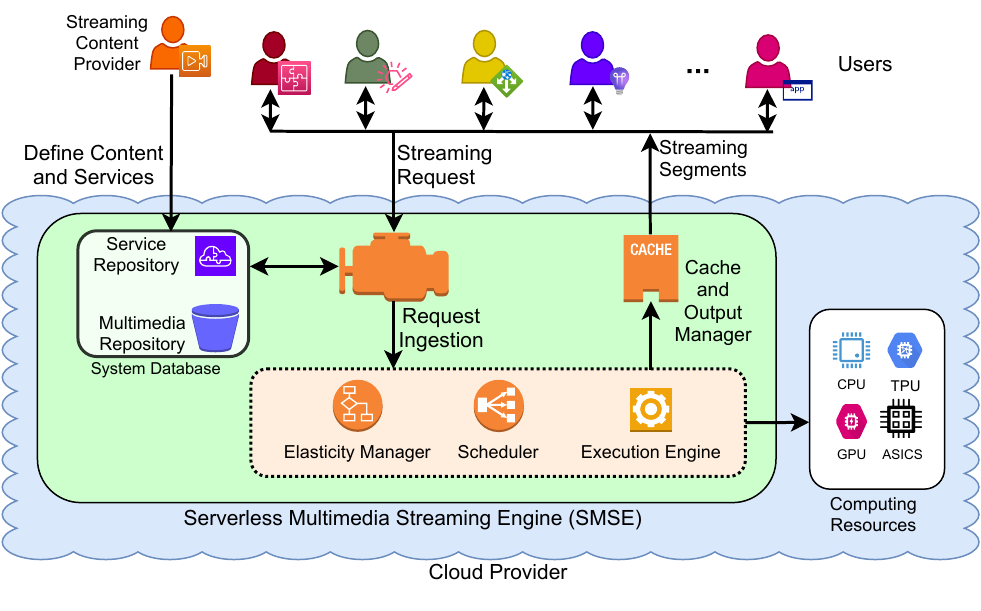}
  \end{center}
  \caption{\small{A bird\revised{'s}-eye view of the \Fname (\name) that serves both the streaming provider (who supplies the multimedia content and define services on them) and different application end-users (who call the services with their streaming requests). \name~can seamlessly and efficiently allocate resources to handle users' requests with respect to their Quality of Service (QoS) demands.}}
  \label{fig:ovrview}
\end{figure}

Multimedia streaming is one of the main Internet services and constitutes more than 75\% of the whole Internet traffic~\cite{cdnstat}. Our motivation in this project is a streaming company (streaming service provider) that develops an application to offer cloud-based live streaming services to disabled people (shown in Figure~\ref{fig:ovrview}). To enhance visual recognition of color-blind viewers, they need to develop a near real-time service to increase the contrast of frames' colors. For deaf viewers, they need a service for dynamic multilingual subtitle generation. For blind viewers, they need a service to provide additional audio description \cite{jin2016video}. Another child entertainment application needs to develop a service to dynamically extract harmful and illicit content from the videos and make them child-safe. A healthcare application needs to develop AI-based service to monitor certain metrics in premature birth cases and inform the medical crew upon detecting important signs \cite{prematurebirth}. 
\emph{Our aim is to develop a dedicated platform, called Serverless Multimedia Streaming Engine (\name), that enables rapid development of streaming services and accommodates them in a cost- and QoS-efficient manner}.
As shown in Figure~\ref{fig:ovrview}, \name~can serve as the cloud back-end for various multimedia streaming applications. Unlike current streaming services, content providers (or even end-users) of the \name~are not limited to the predefined services offered by the cloud provider anymore. Instead, \name~democratizes multimedia programming and enables its users to leverage the high-level abstractions provided by \name~and develop the specific functions demanded by their applications. We call this flexible fashion of multimedia streaming as \emph{interactive multimedia streaming}. 

Upon defining a new user function, \name~is in charge of deploying it on its resources in an efficient and seamless manner. For that purpose, \name~is equipped with a function provisioning method that is customized based on the multimedia workload characteristics to reduce cost of the stream provider and improve QoS of the users in terms of their perceived latency. The function provisioning method captures the spatiotemporal and data-dependency characteristics of multimedia function invocations, then dynamically decides whether to provision a durable or an ephemeral function container while considering the function invocation frequency, its startup latency, and the available system memory. In summary, the contributions of this research are as follows:


\begin{itemize}
\item Developing the architecture of serverless multimedia streaming cloud engine (\name) that enables rapid multimedia service development and accommodates the services in a cost- and QoS-efficient manner.
\item Developing a multimedia function provisioning method that determines the function container allocation based on the spatiotemporal and data dependency characteristics of the functions. 
\item Implementing the prototype of \name~and analyzing its performance under real-world settings.
\end{itemize}

The rest of the paper is organized as follows: 
Section~\ref{sec:bkgd} highlights the related topics of Serverless computing and interactive multimedia streaming. Then, Section~\ref{sec:overview} considers variety of design choices to develop the \name~platform. In Section~\ref{sec:elas}, we explain the efficient multimedia function provisioning method. Next, the framework and function provisioning method are evaluated in Section~\ref{section:platform_eval}. Finally, the paper is concluded in Section~\ref{sec:conclsn}.
\section{Background and Prior Works}\label{sec:bkgd}

\subsection{Serverless Computing in Practice}
Serverless is a misnomer for a specific cloud computing model that aims to provide the illusion of user's tasks being completed without the user noticing any server management details~\cite{mampage2022holistic}. In these systems, all the task profiling and scheduling are handled by the platform automatically~\cite{li2022serverless}. The common practice to utilize serverless computing is to break the user monolithic application into multiple micro-service~\cite{lloyd2018serverless} functions. Each user (in this case, the \ssp) provides the executable functions and the conditions to trigger them (\eg based on a timer or upon arrival of a request). Once triggered, the \textit{task requests} are created in form of ephemeral containerized functions that have to complete processing of the input data within a deadline. From the provider's perspective, a common scheme to efficiently utilize cloud resources is based on a shared queue of arriving tasks with a scheduler that allocates these tasks to an elastic pool of computing resources. 


While provisioning ephemeral function containers is common practice in the serverless systems, prior works (\eg ~\cite{denninnartefficiency23,pu2019shuffling}) reason that it is not necessarily the best practice for all applications.
One reason is that, the startup latency of tasks becomes highly inconsistent with the ephemeral container~\cite{baresi2017empowering}. 
The impact of such inconsistency on the accuracy of the function profiling and task scheduling in the serverless systems have been studied in \cite{shahrad2020serverless,wu2020descriptive,kannan2019grandslam,gunasekaran2020fifer}.
Some research works have been undertaken to remedy the overhead problem and decrease the startup latency of the ephemeral containers~\cite{perez2018serverless,wang2020accelerating}, whereas, some others have tried to avoid it via maintaining the container in memory for a short time after the task execution~\cite{eismann2020serverless}. 
Another reason that is reported as the downside of ephemeral containers in some prior works (\eg ~\cite{sreekanti2020cloudburst,schleier2020faas}) is pertain to their statelessness across consecutive task executions that for applications like big data leads to frequent load/store data between the container and the external storage which can cause up to 500$\times$ slow down in compare to the IaaS clouds~\cite{pu2019shuffling}. Making use of durable containers that stay in-memory across multiple tasks can mitigate such slow down.

\subsection{Multimedia Streaming}

Multimedia streaming is carried out on a set of media \emph{segments} that can be independently processed \cite{bg_2}. Each segment goes through a workflow of processes, before dispatching to the viewer's device. Because of the large size of multimedia contents, first, they are compressed (encoded) based on a standard (a.k.a. codec) \cite{ffmpeg16} into a specific resolution, bit rate, and frame rate. To support heterogeneous display devices, the encoded contents are converted (a.k.a. transcoded) to different formats \cite{liperformanceanalysis,li2016vlsc}. Second, in the transmuxing process, the segments are packaged to enable time-based presentation \cite{Mekuria16} and transmission over the Internet. Third, delivery techniques (\eg HLS and MPEG-DASH) \cite{hlsDash} are used to adapt the stream quality to the network condition. Lastly, to reduce latency, Content Delivery Network (CDN) technology \cite{cdnvideo15} is used to cache streams near viewers. Accordingly, the viewer's \emph{QoS} is defined as the ability to stream each segment within its allowed latency time to create an uninterrupted streaming experience. 

From streaming provider's perspective, cloud resources are infinitely scalable, and virtually all tasks can meet their deadlines if enough resource are provided to them in a timely manner. However, the main limitation that makes tasks miss their deadlines are: (a) the budget limit set by streaming providers to prevent resource over-provisioning that reduces their revenue; and (b) the latency of resource acquisition. Therefore, a large body of research studies have been undertaken to maintain the desired media streaming QoS (\eg~\cite{cvss,li2018cost}) through efficient use of cloud services withing a budget (and resource) limitation. Particularly, the earlier studies have shown that the access pattern of video streams follows a long-tail distribution~\cite{darwich16}. That is, only a small percent of contents (approximately 5\%) are streamed frequently (known as hot streams) \cite{darwich2019cost}. Hence, majority of the video and function services can be put into a deep sleep without any major impact on the overall QoS.

While conventional media streaming can be achieved by preparing a few limited configurations of the content for the viewer to select, interactive media streaming allows users to state their exact specification or even develop that in form of a new service. For instance, a user can stream a video with color correction, object, or face detection services activated.

The conventional approach of streaming through pre-processing all the multimedia segments ahead of time is not feasible for interactive media streaming for two reasons: (A) In live streaming, media are not available for pre-processing; (B) Even for the on-demand streaming, given the long-tail access pattern to the contents and the ability to stream the contents with a combination of interactive services, it is cost-prohibitive to pre-process all the contents for all the services. For instance, just to cover most common display standards, for each media stream, some 90 to 270 versions must be pre-processed and stored~\cite{veillon2019f,liperformanceanalysis}. In fact, pre-processing of streams is performed only for common and frequently-used interactive services and the rest should be processed in an on-demand (\ie lazy) manner~\cite{erfanian2021lwte}. Such processing has to be performed within the latency constraints to guarantee the QoS demands of the users. In this situation, serverless computing is helpful to mitigate the resource allocation burden for multimedia processing.

\section{\Fname (\name)}\label{sec:overview}

The \name~platform can be considered as a domain-specific cloud platform that serves two types of users, namely \emph{\ssp~} and \emph{end-user}. The \ssp s are those who host their multimedia contents on the cloud. \name~offers the \ssp~with a framework to serve multimedia streaming desires in a transparent (\ie serverless) and efficient manner. Unlike conventional SaaS platforms for video streaming, \name~allows \ssp~to select, customize or develop new services (functions) to process the multimedia contents before delivering them to the end-users.  End-users are clients of the \ssp s and request to stream the contents. It is also possible that, upon the \ssp's permission, the end-users can generate (live or upload) contents and cherry-pick service functions for their needs.



\begin{figure*}
      \vspace{-10pt}
 \begin{center}
    \includegraphics[width=0.6\textwidth]{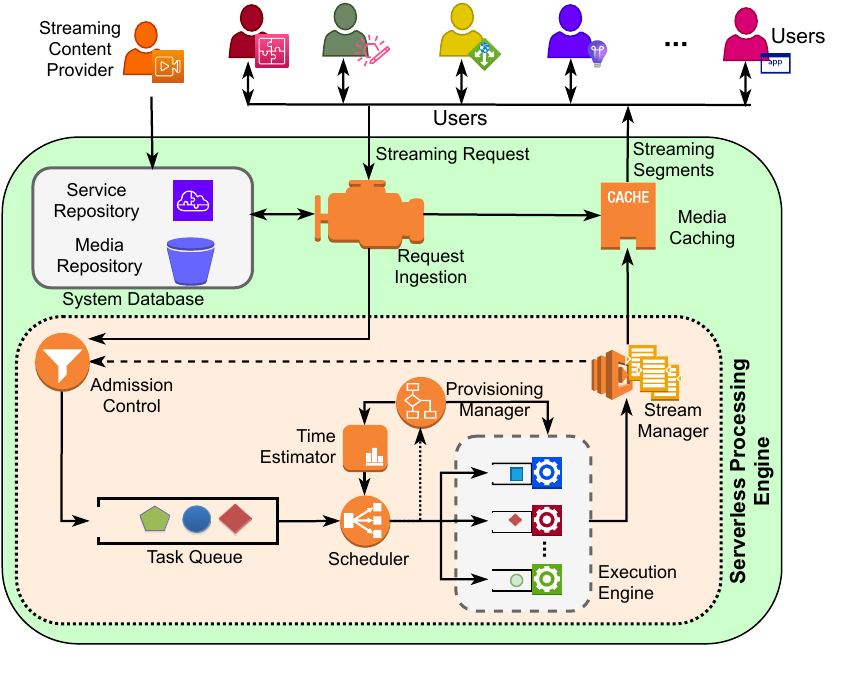}
  \end{center}
  \caption{\small{Architectural view of the Interactive Serverless Media Streaming Engine (\name). The engine deals with the requests of stream providers and end-users. The service providers procure the content and functions to be applied on them. End-users can apply the defined services/functions while streaming the contents.}}
      \vspace{-3pt}	\label{fig:mid}
\end{figure*}

\subsection{Characteristics of Multimedia Streaming Tasks} \label{subsection:function_req}
The following specifications are common across most of the multimedia processing functions (services) and can benefit from a specifically-designed serverless platform over a general-purpose one:

\begin{enumerate}
    \item Each multimedia streaming request generates multiple \emph{independent} tasks and each task is tied to a segment.
    \item Most of the segments are cache-able, yet the content variability is too high to cache all the contents.
    \item The early segments of each stream has a high urgency, because they are required shortly after the request. The urgency of later segments are lowered as the system has more time to process them prior to the presentation time.
    \item Multimedia processing functions in addition to the micro-service code, often have to include other frameworks and libraries (\eg FFmpeg \cite{ffmpeg16} for transcoding and TensorFlow for machine learning services \cite{abadi2016tensorflow}).
\end{enumerate}
    




\subsection{Architectural Overview of \name} \label{subsection:platform_arch} 
Figure~\ref{fig:mid} depicts the architectural overview of \name. In the figure,
the Request Ingestion module is in charge of receiving streaming requests, identifying the possibility of serving the request via the cached contents, and pulling the corresponding segments and functions from the repositories. 
Operating as a serverless computing platform, \name~has a shared scheduling queue to serve tasks from multiple users on the shared resource pool. Other components of the system are the Admission Control, Time Estimator, Scheduler, Stream Manager, Provisioning Manager, and Stream Caching. 

In the rest of this section, we elaborate on each one of these components. The \name~platform has been implemented based on this architecture and it is available in the HPCC lab Github page\footnote{\revised{The prototype implementation of this project is openly available on GitHub at:}\\ \url{https://github.com/hpcclab/CVSS_impl}}.

\subsection{Multimedia Repository}
Multimedia Repository contains segments of the multimedia contents that are available on the platform and streamed upon users' demands. The segments are usually stored in a high-definition format that is suitable for processing by different functions. Upon receiving a streaming request, the repository supplies Request Ingestion with the segments metadata. In addition to the segment's formatting information (\eg resolution, bit-rate, and codec), the metadata can include other data fields, such as those for content rating and age restriction. 

The segment's metadata is used by the Scheduler module to make informed scheduling decisions. Once the scheduler assigns the segment to a particular Execution Unit, a copy of the segments is transferred to that Execution Unit for processing. 

\subsection{Service Repository}
Service repository includes the set of serverless functions that \name~can perform on the multimedia contents. These services are either builtin functions provided by \name, or new stream processing services developed by the stream providers (or even the end-users). 

Conventionally, each serverless function is a stateless and standalone unit that is executed within an \emph{ephemeral container}. That is, the function is initiated to execute one task within a container and then it terminates. This is also known as \emph{cold start} function provisioning. However, due to the nature of multimedia processing that has a large framework dependency footprint, \name~also supports frequently-used functions to be executed within a \emph{durable container}, which is also known as \emph{warm start} function provisioning. In this manner, the durable container remains active (in-memory) after processing one task to serve other tasks of the same function. This enables \name~to significantly reduce the overhead of loading large functions that are frequently invoked. 


\subsection{Request Ingestion}
Request Ingestion handles all the processing requests being made by end-users. Upon arrival of a streaming request, this component authenticates the user accessibility to the content (\eg checking the user billing balance). Then, for each streaming request, it generates multiple tasks that can be processed concurrently. In essence, the Request Ingestion component converts the user request into tasks with individual deadlines (based on the the segment's presentation time) that can be handled by the serverless processing engine of \name. In the event that the requested content is already cached, Request Ingestion notifies the Caching component to bypass the processing and directly stream the content to the user. 



\subsection{Task Admission Control} 
Admission Control is the front gate of the Task Queue where tasks wait to be mapped by the scheduler. The Admission Control is in charge of assigning a priority to each segment based on its urgency. Furthermore, it can be extended to perform task deduplication (a.k.a. task merging) on the arriving identical or similar tasks to make use of the cloud resource more efficiently, as elaborated in our prior study~\cite{tpdschavit}.

\subsection{Task Queue}
Each \emph{task} in \name~comprises the following items: 
\begin{itemize}
    \item An independent segment of the requested stream and its metadata. Let $x$ denote the stream identifier and $y$ denote the segment identifier, then stream $S_x$ is defined as a set of segments $s_{xy}$, where $S$ represents the set of all multimedia streams.
    \item Function $i$, denoted $f_i$, that is available in the Service Repository $F$ (\ie we have $f_i\in F$).
\end{itemize}

Accordingly, a task in \name~is defined as ($s_{xy},f_i$) pair and the task queue (denoted $T_q$) is a set that is formally defined as $T_q=\{$($s_{xy},f_i$)$| $($s_{xy},f_i$)$\in S\times F\}$. 
A task undergoes four states to be scheduled, namely \emph{Unmapped:} the task is yet to be assigned to a computing unit; \emph{Pending:} task is assigned and waiting to be executed on a particular computing unit;  \emph{Running:} task is being executed on the assigned computing unit; and \emph{Completed:} task has completed its execution on the computing unit. Only tasks in the unmapped stage reside in the Task Queue and others are already dispatched to a specific computing unit.


\subsection{Task Scheduler}
In consultation with the Time Estimator component, Scheduler maps each task to an existing execution unit. We adopt the mapping method developed in our prior works \cite{li2018cost,li2016high} to schedule video streaming segments on cloud resources with the aim of maximizing the user QoS via minimizing the startup latency of the requested stream. It is a priority-based scheduler that, for a given stream, it assigns a higher precedence to the first few segments over the later ones. This behavior results in a short startup latency and maintains the fairness across multiple users. We adapted the scheduling methods for the \name~scheduler as follows: First, not all Execution Units can process any given task type. This is due to the nature of heterogeneous task and computing resources along with service container availability. Second, even if the Execution Unit is capable of processing the task type, it can be oversubscribed (\ie its waiting queue gets too long), hence, it is ruled out from the list of potential assignment candidates. 

We note that \name~has been designed to get its scheduling method as a modular, plug and play component. In the prototype implementation, in addition to the aforementioned method, we have provided several other baseline scheduling methods that can be opted in the configurations for testing purposes. Users of \name~can extend and examine the scheduling methods based on their specific requirements. For instance, scheduling methods can be developed to comply with the specific billing policies defined by the \ssp~in a given multimedia cloud system.




\subsection{Task Execution Time Estimator}
Time Estimator provides a matrix where each entry represents the estimated execution (and completion) time of a task type (\ie a function) on each Execution Unit (\ie machine type) \cite{liperformanceanalysis}. The matrix entries are used by the Scheduler to make informed mapping decisions and improve the efficacy of the \name~platform. 


The module is designed to be extensible and users of \name~can override the default estimation method (\eg using machine learning-based methods). Currently, we have prototyped two Time Estimator methods for \name. In one implementation, called profile-mode, the Time Estimator reads the predefined matrix of the expected execution time of each function on each Execution Unit type. This mode is  deterministic and is primary suitable for testing and simulating tasks on the scheduler. In another implementation, called learning-mode, Time Estimator accumulates historical data from prior task executions to form the knowledge for estimating the execution time of each task type in the later occurrences. In this mode, Time Estimator does not discriminate the estimation data between different media segments. In other words, a task that involves the same operation on the same Execution Unit yields the same Time Estimation, regardless of the involved media segments. The learning-mode is suitable when a high amount of task type variation is expected, and there is no complete execution time profile available.

\subsection{Execution Engine}
\label{subsec:PU}
Execution Engine consists of a scalable pool of Execution Units. We define a Execution Unit as the most granular computing resource in the view of the scheduler that can have its own local queue (a.k.a. machine queue) that the scheduler can assign the tasks to. \name~can be flexibly configured to work with bare-metal, Virtual Machine (VM), or container as its Execution Unit. However, as we explain in the evaluation section, containers impose a relatively low overhead and offer task isolation, thus, in the rest of this paper, the default Execution Unit is deemed to be containers. 

Once started, a Execution Unit listens to its machine queue (implemented via RabbitMQ \cite{dobbelaere2017kafka})
for the tasks assigned by the scheduler. The machine queue is vital for efficient usage of computing resources via minimizing the idle times between the scheduling events. Upon a task arrival to the machine queue, the Execution Unit fetches the corresponding segment prior to its execution. For the tasks with normal priority, machine queues are handled in the first-come-first-serve (FCFS) manner. However, high-priority tasks skip the queue for urgent execution.



\subsection{Function Provisioning Manager} 
Provisioning Manager is a critical component in \name~that is responsible for allocating resources to the functions. This responsibility can be divided into two conjunctive parts, namely elasticity manager and container provisioning. Elasticity manager handles the scale in and scale out of the computing resources based on the users' demand. That is, it allocates more cloud resources (in form of bare-metal of VM) upon surge in the workload and vice versa. However, the container provisioning manager is more granular, and it deals with the efficient provisioning of functions containers on the allocated computing resources. 

As mentioned earlier, in a container-based serverless platform, a function is provisioned either using ephemeral containers that are loaded from the storage for each invocation; or durable containers that remain in the memory to serve the next invocations of the same function.
The downside of the ephemeral containers is the nontrivial time overhead they impose to load the container image \cite{lloyd2018serverless}, whereas, the downside of durable containers is the memory space consumption that prevents it to be applied to all functions in the serverless system. As such, container provisioning manager is in charge of dynamically determining if the container is to be provisioned as a ephemeral or durable container. In the latter case, it also determines the number of needed warm container instances, such that the upcoming tasks of that function can be served with respect to the users' QoS demands. The way Provisioning Manager operates is decisive on the performance we can gain from the allocated cloud resources. Therefore, we dedicate Section~\ref{sec:elas} to propose an efficient container provisioning method for \name.  


%



\subsection{Stream Manager}
Stream Manager resides near the end of the processing pipeline. It keeps track of the requested segments of all streams and assures all the segments are processed in a timely manner. In the event of missing a segment, Stream Manager requests the Admission Control to resend the missing segment with an urgent priority. Stream Manager also enables the multi-stage tasks (\ie workflows) in \name. When the Stream Manager detects a task that needs to be processed further with a function to satisfy the viewer's expectation, it generates the new task accordingly.

\subsection{Multimedia Caching}
The last stage before the content is transferred to the viewer's device is the Multimedia Caching. It manages the streaming channel (via direct streaming or CDN). The caching module determines the hotness (\ie popularity) at the segment granularity, as discussed in our prior study~\cite{darwich2019cost}. That is, the segments (tasks) that are anticipated to be streamed again in the near future are cached for reusing in a local caching server or on the CDN~\cite{veillon2019f}.

\begin{figure}
 \begin{center}
    \includegraphics[width=0.5\textwidth]{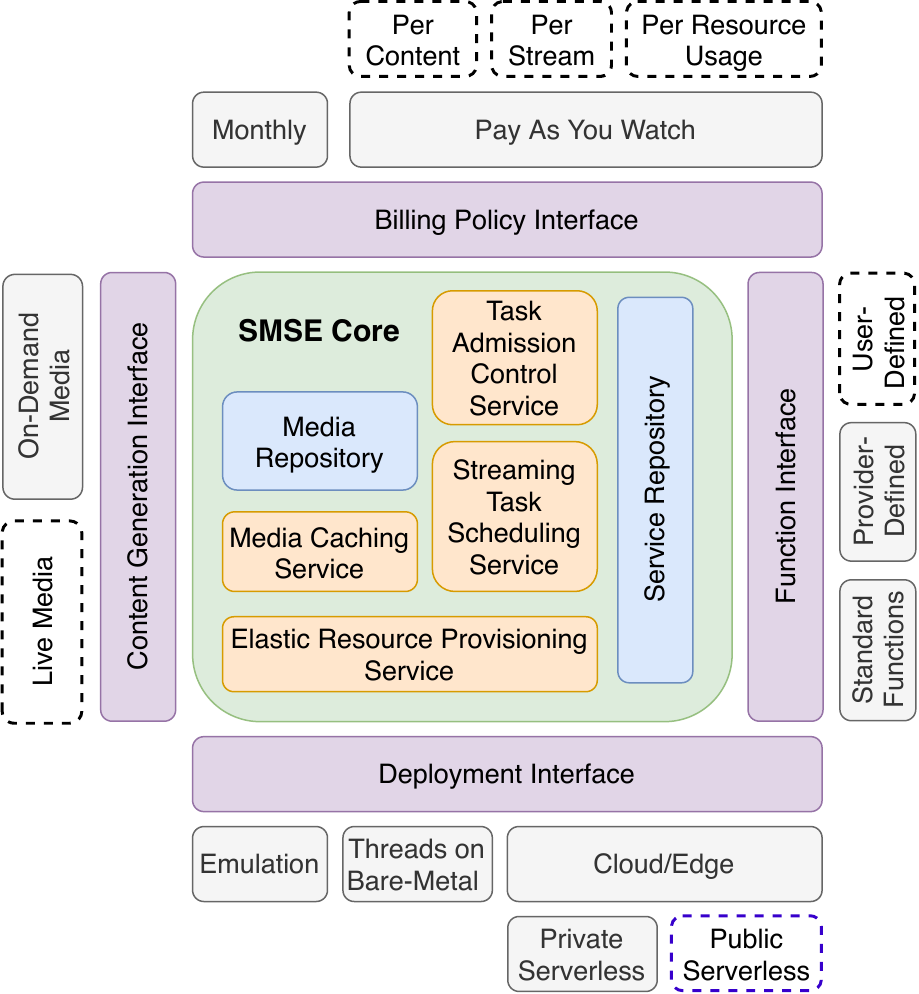}
  \end{center}
  \caption{\small{Expandable Interfaces in Serverless Media Stream Processing Engine (\name). Purple boxes outside of the \name~Core represent different interfaces that collectively make the system flexible. Dashed-line boxes indicate planned components that yet to be developed in the prototype. }}
      \vspace{-3pt}	\label{fig:class}
\end{figure}

\subsection{Configurability and Extensibility of \name}
\name~has two deployment modes. The main one is the full stack multimedia processing platform that serves user requests on the cloud resources. The other one is the emulation-mode that is designed to help researchers and practitioners to examine the performance of different methods and components in the system without the need to utilizing real cloud resources. In this mode, \name~is transformed to a discrete event emulator that enables comprehensive studying of the behaviors of different methods under various workload scenarios and within a limited time. In the emulation mode, User requests are pre-generated by Ingestion Control and arrive at a certain time stamp. The execution time of each task is sampled within a range dictated by the Time estimator component. The emulation logs generated in this mode represent the deadline miss rate of the tasks (segments) of different streams along with other relevant performance metrics. 

Figure~\ref{fig:class} depicts the developed and planned components of the \name~platform. At the core, we can see the main components of \name~that were described in the previous parts. The core is surrounded by four interfaces that allow new components plugged into the system. The Deployment Interface defines the functionalities that must be implemented on each underlying platform. Currently, \name~supports three platforms that are shown with solid lines. The Function Interface is for service developers and it works directly with the Service Repository. The Content Generation Interface is to support different forms of streaming, and finally the Billing Interface offers the flexibility to the \ssp s to develop their custom billing policy. Allowing the \name~to serve multiple use cases, \name~saves configuration parameters in a XML-based configuration file. Any custom-made component can also define its options in the same configuration file. \revised{These flexible interfaces help the \name~keep up with upcoming advancements~\cite{mampage2022holistic, li2022serverless} in cloud and serverless~\cite{bruno2022graalvisor} computing}.

\subsection{Prototype's Limitation and Room for Future Improvements}
\begin{itemize}
\item \revised{The scheduler assumes that Task Execution Time Estimator can estimate the task execution time. This is not suitable in the circumstances where tasks' execution times do not follow any pattern.}
\item \revised{Scheduling each and every task without a scheduler bypass can be too granular. A Multi-level scheduler which schedules in two levels (stream and task) can reduce the scheduling latency for the latency-sensitive or fall-behind video streams.}
\item \revised{The prototype Streaming Task Scheduling Service is not designed for scalability. The scheduler itself can become the scale out bottleneck. In the future, a distributed scheduler can be developed to resolve this bottleneck.}
\end{itemize}
\section{Function Provisioning in \name}
\label{sec:elas}
\subsection{Problem Statement}
\begin{figure}[ht]
 \begin{center}
    \includegraphics[width=0.55\textwidth]{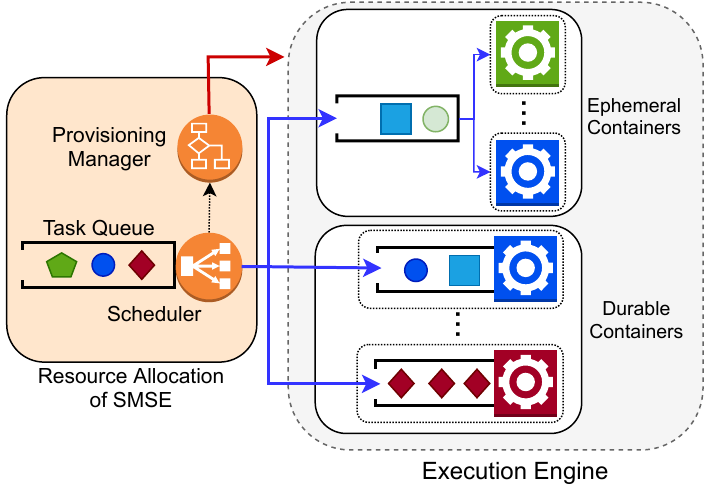}
  \end{center}
  \caption{\small{An example of the Provisioning Manager module controlling function containers in two configurations. The top one has a centralized machine queue and creates one ephemeral container for each task. The bottom configuration shows in-memory durable containers for different functions (task types) where each container has a dedicated local queue that is accessible to the task Scheduler.}}
    	\label{fig:resources}
\end{figure}

Provisioning ephemeral containers allows multiple functions to share the computing resources (see Figure~\ref{fig:resources}), hence, it is an effective way to support a large function repository within a limited memory space. However, ephemeral containers impose a significant overhead due to repeatedly launching and terminating the containers. This overhead comprises the time for: (a) fetching the container image from the repository; (b) starting the container; and (c) initializing the task data (multimedia segment) from the shared queue into the container. In contrast, a durable container stays in the memory and has a task queue that is directly accessible to the task scheduler (see Figure~\ref{fig:resources}) and can serve multiple tasks of the same function. In compare to the ephemeral containers, durable ones can mitigate the overhead via eliminating repeatedly launching and terminating containers, hence, reducing the task execution latency.

Public serverless computing platforms have to maintain a large repository of service functions to serve their users. To minimize the latency of the functions, ideally, the functions should be provisioned using durable containers (\ie warm start). However, due to the memory space limit of the hosts, it is not affordable and cost-efficient to allocate durable containers for all functions. This is reasonable, particularly when we know that the usage (invocation) pattern of the functions follows a long-tail distribution, \ie only a small portion of the functions invoked frequently and the rest are rarely accessed \cite{shahrad2020serverless}. These factors make the management of memory space across functions a challenge in \name~and other FaaS-based systems.

The challenge is how to determine which functions should be durable and which ones remain ephemeral? Unarguably, the whole memory space cannot be consumed for durable containers and a portion of it has to be reserved to handle the execution of ephemeral containers. As such, the follow up challenge is how much memory space has to be reserved for ephemeral functions? And, for durable containers, the challenge is how to determine the number of container instances to be made durable? These challenges are inter-related and cannot be studied in isolation. For a given function, allocating more durable containers consumes the memory space that can otherwise be allocated to make other functions durable. 

As mentioned earlier, a task is defined in the form of a segment-function pair. Provided that each requested stream comprises several segments, the task arrival pattern in \name~have a spatiotemporal property. That is, upon receiving a streaming request, the upcoming set of tasks are different segments of the stream that are processed with the same function. Knowing the spatiotemporal property of invocations and the dynamic popularity of functions, the Provisioning Manager module periodically monitors the functions and determine the provisioning strategy for each function such that eventually more tasks can be completed using the available cloud resources. 

\subsection{Proposed Function Provisioning Method}
The goal of this method is to efficiently utilize the memory space of cloud resources via allocating them to function containers, such that the wasted overhead time is replaced with the useful time to execute tasks, thereby, improving the user experience without requiring more cloud resource. Finding the optimal function provisioning is a bin-packing problem that is known to be an NP-hard problem. Therefore, we propose a heuristic method that sorts the tasks based on the \emph{benefit per cost ratio} of provisioning a function container.



Let $F$ be a function repository (set) in form of 
$F=\{f_1, f_2,..., f_k \; | \; k=Repository\;Size \}$. Tasks of function $f_i$ are served by one or more coexisting (\ie concurrent) instance(s) of the function container, represented as $f_{i1}$ to $f_{in}$, where $n$ is the maximum concurrency degree of $f_i$. The maximum concurrency degree of function $f_i$ is the highest number of containers instantiated concurrently to meet the deadlines of tasks of $f_i$ during last $\alpha$ provisioning events (called history window). For instance, assume that we set the last three time slots as the history window (\ie $\alpha=3$) and the maximum number of concurrent containers for $f_i$ during this window has been four, thus, the maximum concurrency degree of $f_i$ is considered four. 

For the function container $f_{ij}$, we define its \emph{utilization} ratio, denoted $U(f_{ij})$, as the time container has been utilized to execute tasks to the total time window. The utilization value spans between 0 (always idle) to 1 (always utilized). For container $f_{ij}$, Provision Manager probes its utilization, and the number of times it is used to run a task (a.k.a. triggering frequency), denoted $\mu(f_{ij})$ during the history window. For instance, if function container $f_i1$ is triggered 10 times, and it is idle for five percent of its time during its history window, then we have $\mu(f_{i1})=10$ and $U(f_{i1})=0.95$. Meanwhile, if function container $f_{i2}$ is triggered 2 times with utilization of 20\%, then $\mu(f_{i2})=2$ and $U(f_{i2})=0.2$. 

Let the end-to-end execution (a.k.a. turnaround) time of a task of $f_{i}$ on an ephemeral and a durable container on the same host be $s1$ and $s2$ seconds, respectively. Then, we define the \emph{durable time saving} as $\delta({f_i}) = s1 - s2$ and we have $min(\delta({f_i}))=0$. For function container $f_{ij}$, we estimate the \emph{benefit} of provisioning it as a durable container, denoted $B(f_{ij})$ based on Equation~\ref{eq:benefit}.

\vspace{-10px}
\begin{equation}
B(f_{ij}) = \mu{(f_{ij})} \cdotp \delta({f_i})
\label{eq:benefit}
\end{equation}

Intuitively, ephemeral containers require no dedicated memory allocation. However, a closer investigation reveals that such functions do require memory upon invocation, hence, a portion of memory space must be reserved for the execution of ephemeral functions and we cannot allocate all the available memory to the durable containers. Let $M(f_i)$ represent the actual memory consumption of a durable container for $f_i$ during the history window. However, the average memory usage of function container $f_{ij}$ throughout the history window, denoted $M_r(f_{ij})$, is calculated by considering both the memory requirement of $f_i$ and the container utilization within the history window, as shown in Equation~\ref{eq:reservemem}. 
\vspace{-10px}
\begin{equation}
M_r(f_{ij}) = M(f_i)\cdotp U(f_{ij}) 
\label{eq:reservemem}
\end{equation}
\vspace{-5px}

As such, the extra cost of allocating container $f_{ij}$ as a durable one, denoted $C(f_{ij})$, is calculated based on the difference between $M(f_i)$ and $M_r(f_{ij})$, as shown in Equation~\ref{eq:cost}.
\vspace{-10px}

\begin{equation}
C(f_{ij}) = M(f_i) - M_r(f_{ij})
\label{eq:cost}
\end{equation}

Then, the Benefit per Cost Ratio of having function container $f_{ij}$ as durable is calculated based on the following equation: $BCR(f_{ij}) = {B(f_{ij})}/{ C(f_{ij})}$.

In this equation, we notice that when function container $f_{ij}$ is highly utilized (\ie $U(f_{ij})\rightarrow 1$), we have $M_r(f_{ij}) \rightarrow M(f_i)$, therefore, $C(f_{ij}) \rightarrow 0$ and $BCR(f_{ij}) \rightarrow \infty$. Conversely, a rarely-invoked function (\ie $U(f_{ij}) \rightarrow 0$) leads to  $BCR(f_{ij})\rightarrow 0$. From another perspective, we notice that when a durable container yields a high time saving (\ie a high $\delta$ value), it leads to a high $BCR$ value as well and vice versa.

We leverage this analysis to develop the function provisioning method, shown in Algorithm~\ref{fig:allocateFn}. This method is invoked periodically, at each provisioning event, to determine the set of durable containers for the next provisioning time slot. 
The input to this algorithm is the function repository ($F$), memory information, and the set of $BCR$ values for all function containers.
Let $A$ a vector of pairs that maintains the number of durable containers for each function in the repository. In Line 1, $A$ is initialized to zero for all functions. Let $M_A$ represent the total memory allocated for durable functions and it is initialized to 0 in Line 2. Next, in Line 3, $M_R$ contains the reserved memory space for the ephemeral function containers according to Equation~\ref{eq:reservemem}. Then, in Lines 4---9, durable functions are allocated as long as there is available memory and there is a function container that can benefit from durable provisioning. In particular, Line 5 selects the function container with the highest benefit per cost ratio ($BCR$) and its corresponding durable allocation is incremented in Line 6. The allocated memory for durable function and the reserved memory for ephemeral functions are adjusted accordingly in line 7 and 8. Finally, the resulting allocation set is returned in Line 10.


\begin{algorithm}[t]
    \caption{Pseudo-code of the function provisioning method.}
    \label{fig:allocateFn}
    \begin{algorithmic}[1]
         \REQUIRE
            $F, M, M_r, BCR $ as defined in Section~\ref{sec:elas}; $M_T$ which is the total usable memory of the host;

            

                      
            
             

         \ENSURE
             $A$: the set of function containers that should be allocated as durable containers;
        \STATE $A = \{ ({f_i,0) | f_i \in F}\}$
        \STATE $M_A \leftarrow$ 0
        \STATE $M_R \leftarrow \displaystyle\sum_{\forall f_i \in F}\sum_{\forall j}
        [M_r(f_{ij})
        ]$
        

        \WHILE{$M_T-M_R-M_A > 0 $ AND  $ \displaystyle\max_{\forall i, \forall j}(BCR(f_{ij}))>0$} 
        \STATE  $f_{max} \leftarrow f_{ij}$ that has the maximum BCR value
        
        \STATE Update $(f_{max}, c_i) \in A$ to $ (f_{max},c_i +1)\in A$
        
        \STATE $M_A \leftarrow M_A + M(f_{max})$
        \STATE $M_R \leftarrow M_R - M_r(f_{max})$
        \ENDWHILE
        
        \STATE \textbf{Return $A$}
    \end{algorithmic}
\end{algorithm}

\section{Performance Evaluation}
\label{section:platform_eval}

We developed the prototype of the \name ~platform. In this prototype, we particularly focus on the multimedia processing engine and its scheduler. The platform enables us to a) verify the suitability of containerization to offer user-defined multimedia processing functions. We note that this has not been the case in multiple other contexts (\eg \cite{ghatrehsamani2020art}); b) verify that the notion of Durable Functions can be effective in reducing the execution time and the potential gain is not negligible; c) evaluating the scheduling performance of the proposed task provisioning method against two common approaches that are scheduling durable function statically, and not utilizing durable functions at all.

\subsection{Experimental setup.}

The multimedia repository we used for the evaluation includes a set of video files that are diverse both in terms of their content types and their length, which is in the range of [10, 220] seconds. These video files are split into 5---110 two-second-long video segments. These benchmark videos are publicly available for reproducibility purposes\footnote{The benchmark video dataset used in the experiments \revised{is openly available on GitHub at}:\\ \url{https://github.com/hpcclab/videostreamingBenchmark}}. 

The prototype implementation of \name~is provided with 16 sample functions in its service repository. All of these functions have separate container images. Each function has one or more input parameters that can be configured. For example, the \texttt{frame\_rate\_change()} function has an input parameter that can be set to 30 or 24 that dictates the frame rate of the output video. 
To avoid any randomness in the result, we repeat each experiment 10 times and report the mean and 95\% confidence interval of the results.


\subsection{Evaluating the Startup Latency of Execution Units}
The goal of this experiment is to compare the startup latency (overhead) imposed by various forms of Execution Unit supported by \name, namely bare-metal, VM, and container. For this experiment, we choose one function and add libraries (FFMpeg and TensorFlow) to it to create function packages in three different sizes. We assured that all the three function packages read (load) the libraries into the memory before beginning the task execution. Then, we run the task for 10 times and measure the startup latency in each case. 

\begin{figure}[ht]
	\centering
	\includegraphics[width=0.4\textwidth]{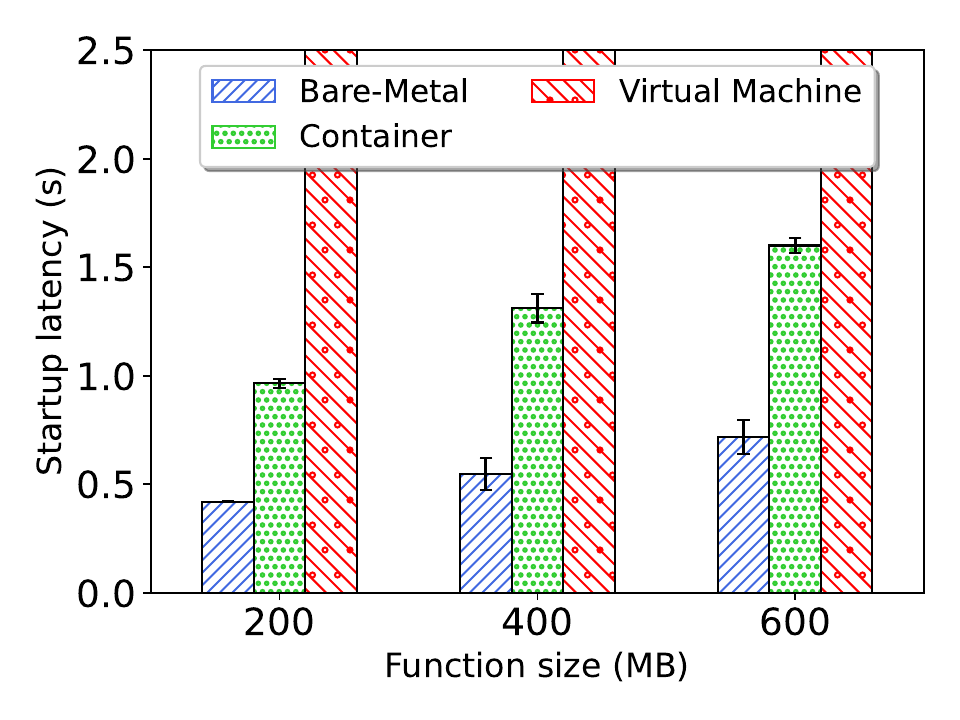} 
	\caption{Comparing the startup latency (overhead) of three Task Execution Unit deployments, namely bare-metal,  container, and Virtual Machine (VM) \revised{with 3 function sizes}. The vertical axis shows the time (in seconds) to launch the Execution Units.}
	\label{fig:StartUpLatency}%
\end{figure}

Figure~\ref{fig:StartUpLatency} shows the startup latency of different forms of the Execution Unit deployments (in seconds). The latency of launching \revised{Qemu 2.11 with Ubuntu} VMs \revised{(from a complete cold state)} in our setup \revised{(Intel Xeon 16 cores with 16 GB memory)} goes over 12 seconds which is unchartable. \revised{It is possible to pause the VM in a state that can go up faster. However, doing so requires significantly more memory to maintain.} Performance-wise, we observe that using bare-metal for the Execution Unit has the lowest startup overhead. However, such configuration lacks isolation and cannot flexibly scale. In particular, lack of isolation causes functions running on a bare-metal machine potentially interfere with each other, \eg due to desiring conflicting software libraries. The figure shows that using container as the Execution Unit eliminates such a downside with a minimal overhead. Therefore, in the rest of experiments we consider containers as the Execution Unit for functions.

\subsection{Evaluating Ephemeral versus Durable Container}
\label{subsec:Evalreusecontainer}
\begin{figure}[ht]
	\centering
	\includegraphics[width=0.4\textwidth]{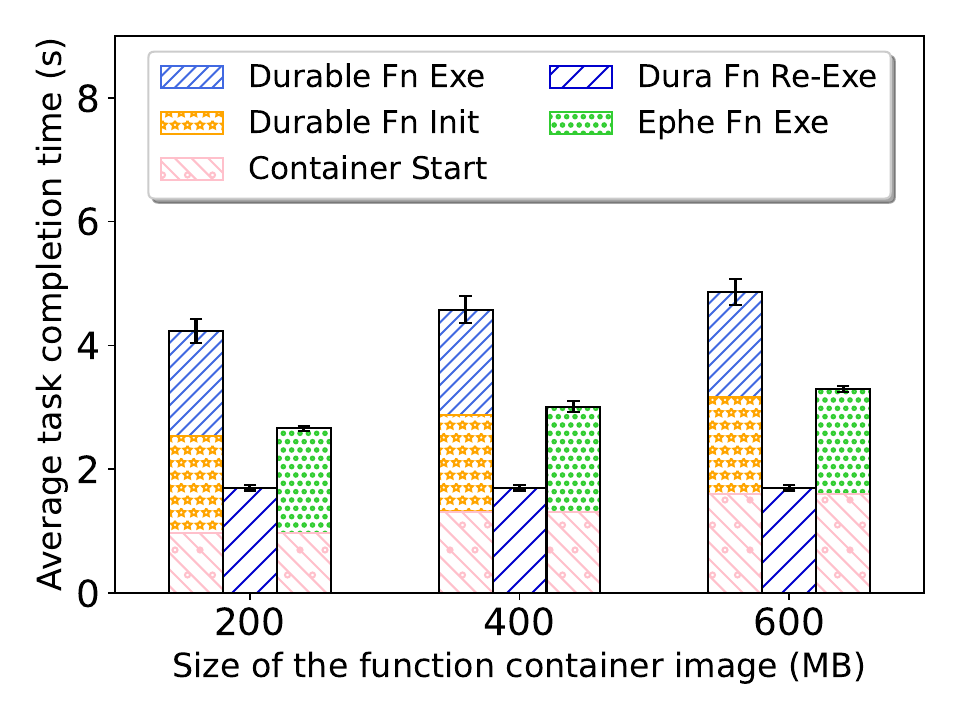} 
	\caption{Comparing the completion time of tasks on function containers with various sizes. The containers are deployed as durable, reused durable, and ephemeral containers. In each case, the contributing factors to the completion time are reported. For durable functions, there are the start, initialization, and execution times; for reusing durable containers there is the task execution time (noted as \texttt{Dura Fn Re-Exe}); and for ephemeral containers there is the container start and execution times (noted as \texttt{Ephe Fn Exe}).}
	\label{fig:ContainerRoundTrip}%
\end{figure}

Knowing that container is a suitable vehicle to provision containers, in this experiment, we plan to dive deep into the detailed overheads imposed by different forms of containers. For that purpose, we evaluate the time overhead of deploying ephemeral versus durable containers for functions. Similar to the previous experiment, we form containerized functions in three different sizes and measure the average time to complete a task in three scenarios: (a) with a durable container allocated for the first time; (b) with an in-memory (reused) durable container; and (c) with an ephemeral container. In each scenario, as shown in Figure~\ref{fig:ContainerRoundTrip}, we granularly report the latency of different factors contributing to the overhead. 
For ephemeral containers, the task completion time contains the time to start the container and the time to execute the function. Durable containers include an additional step to initialize the message queue for the container, as explained in Section~\ref{subsec:PU}. However, once the durable container is initialized, the time for the next tasks assigned to the durable container includes only the data (segment) transfer and execution time latencies. We note that, in our experiments, the transfer time was in the range of [6,30] milliseconds which is negligible (and not plottable) in compare to the task execution time that is in the scale of seconds.

In Figure \ref{fig:ContainerRoundTrip}, we notice that the initialization overhead of the durable container pays off when reusing it for the next tasks of the same function. That is, the time saving from launching a new ephemeral function container versus reusing a durable container can add up and surpass the extra initialization time of the durable container after three repeated invocations. As such, we can conclude that for functions containers that are used by more than three tasks, it is recommended to use durable function containers to minimize overall makespan. According to the figure, the container image size is not a decisive factor on the overhead, nonetheless, the benefit of reusing durable containers becomes more significant for larger containers.

\subsection{Evaluating Scalability of the Provisioning Manager Method}
 
\begin{figure}[ht]
	\centering
	\includegraphics[width=0.4\textwidth]{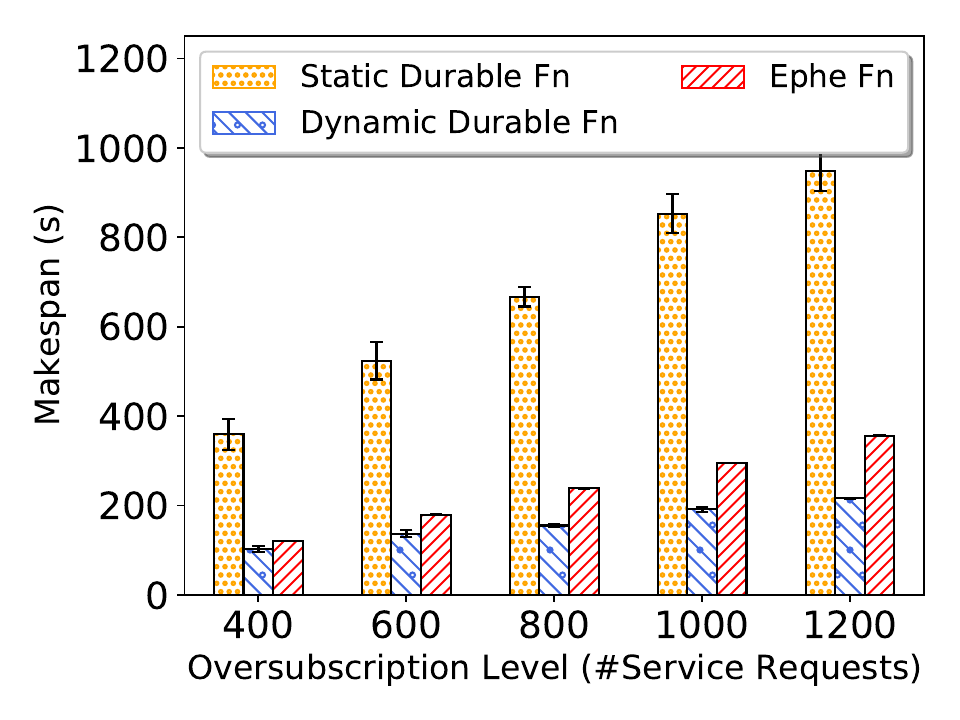} 
	\caption{Comparing the makespan time of completing different number of tasks arriving to \name~within a fixed time interval (known as oversubscription level) using three different provisioning methods (static durable containers, dynamic durable containers, and ephemeral containers).}
	\label{fig:makespan}%
\end{figure}

In this experiment, our aim is to evaluate the impact of Provisioning Manager within the \name~platform. To measure this impact, we study the system under different oversubscription levels with 400---1200 tasks (a.k.a. service requests) arrive to the system within a fixed time interval, as depicted in the horizontal axis of Figure~\ref{fig:makespan}. Tasks arrive in batches of 5---20 consecutive segments with two-second inter-arrival times between tasks of the same set. Each task is a pair of (segment, function). The first segment in a task set is randomly chosen from one of the videos in the benchmark dataset, and then, a sequence of 4--19 following segments are also chosen as the next tasks in the set. All of the tasks in a set use the same function configuration that is chosen randomly from the service repository. We note that at any given time more than one task sets can arrive. To mimic real video steaming workloads, we configure each workload trace to repeatedly toggle the arrival rate of task sets between a base-load period (lull) and a high-load period (peak) with double arrival rate. Each base period is approximately three times longer than the high-load period.

We measure the makespan time (\ie the total time to complete all the tasks) with the following three function provisioning methods: 
\begin{enumerate}
    \item Ephemeral containers (represented as \texttt{Ephemeral Fn}) that treats all functions as cold start. This is a common practice in serverless computing platforms to ensure the maximum resource utilization, because it does not reserve any memory space. 
    \item Static durable containers (represented as \texttt{Static Durable Fn}) where all functions are warm start from the beginning. In this case, the provisioning manager provisions all the functions being examined with durable functions, thus, there is no container startup overhead. 
    \item Dynamic durable containers (represented as \texttt{Dynamic Durable Fn}) where the durable function containers are dynamically adjusted based on the method explained in Section~\ref{sec:elas}.
\end{enumerate}

Figure~\ref{fig:makespan} shows the makespan time to complete all the tasks using the three methods. We observe that, surprisingly, \texttt{Static Durable Fn} leads to a significantly higher makespan time than the \texttt{Ephemeral Fn} method. This is because the arriving task-type distribution is not uniform and the function popularity changes over time, therefore, the static provisioning method consistently over-utilizes certain function containers while under-utilizes the others at the same time. It is this idling of some function containers that causes this substantially longer makespan time. Nevertheless, we observe that the proposed \texttt{Dynamic Durable Fn} yields the lowest makespan time and the difference reaches to 30\% for higher oversubscription levels. The reason is that this method monitors the functions' usages and provisions durable containers accordingly. Thus, it gathers the benefits of low startup overhead, provided by the durable containers, as well as the high resource utilization, provided by the ephemeral containers.

\section{Conclusions and Future Work}\label{sec:conclsn}
In this paper, we introduced and described the prototype implementation of domain-specific serverless cloud platforms and, particularly, \name~that is designed to be highly extensible and democratize cloud-native service development in the context of multimedia streaming. The platform allows \ssp~and even end-users to develop their own functions while offering tools and services for cost- and QoS-efficient allocation of resources to serve multimedia tasks. Within this platform, we particularly developed a method to dynamically provision functions with durable containers. We observed that the it is counterproductive to lavishly deploy durable function containers. This approach not only under-utilizes cloud resources, but it also falls behind ephemeral function containers in terms of the makespan time. However, the dynamic function provisioning method that we developed gains the benefits of both ephemeral containers in terms of resource utilization, and durable containers in terms of imposing a low overhead. In particular, we noticed that the benefits of function provisioning method gets more substantial when the system is oversubscribed and a more efficient use of resources is demanded. 

Although we develop this work the context of multimedia streaming, it is envisaged that the idea of domain-specific cloud platforms will be expanded to other domains, such as machine learning and big data, in the near future. In this case, an application is rapidly developed by the mashup of services from multiple domain-specific serverless clouds. There are also several avenues that we plan to extend \name. One avenue is to enable higher-level programming abstractions, such as Object-as-a-Service, that can mitigate the burden of developing cloud-native applications and help democratizing it. Another avenue is to extend \name~across edge-to-cloud continuum, such that the serverless paradigm is applied seamlessly across the tiers in the continuum. That is, upon invoking a function, it is seamlessly executed on a computing tier that satisfies the cost, privacy, and latency requirements of the end-user.

\section*{Acknowledgments}
We are thankful to anonymous reviewers of this paper. This research is supported by the National Science Foundation (NSF) Computer Network Systems (CNS), CAREER Award\# 2047144. 

\bibliographystyle{plain}

 \bibliography{references}

\end{document}